# Location Update Accuracy in Human Tracking system using Zigbee modules

B.Amutha ,                                    M.Ponnavaikko
Research Scholar,                             Vice Chancellor
SRM University,                               Bharathidasan University
Chennai,Tamil Nadu,India.                     Trichy,Tamil nadu,India

**Abstract:**

**A location and tracking system becomes very important to our future world of pervasive computing. An algorithm for accurate location information is being incorporated in the human walking model and in the blind human walking model. We want to implement an accurate location tracking mechanism using Zigbee along with GPS, we have incorporated Markov chain algorithm for establishing accuracy. Normal Human and blind human walking steps were actually taken in the known environment within our campus and the Markov chain algorithm was used for smoothening the stepwise variation in location updates. A comparison module is also implemented to show the difference between normal human and blind human walking step variations. This accuracy is used for designing a blind tracking device so that the device can be used by the blind for finding the path without obstacles. We present a system-level approach to localizing and tracking Human and blind users on a basis of different sources of location information [GPS + Zigbee]. The system can be applied outdoors especially for avoiding accidents, GPS as the source of location data. Performance evaluation shows that the system is accurate and it is a future path finding device with service for the blind.**

## 1. INTRODUCTION

A system-level approach to the problem of localizing and tracking a human user who assumed to be carrying a device for locating is considered in this paper. The device is GPS along with a Zigbee module which consists of an odometer for stepwise location measurement. This device estimates its location by using GPS as it is incorporated within the device itself. System architecture for a sensor network for human tracking and for the benefit of the blind persons to identify their future path stressed the necessity of the sensor localization.

A Human tracker should track accurately for long sequences; self-start; track the person independent of activity; be robust to drift situations; track multiple people simultaneously; track through occlusions; and be computationally efficient and to provide accurate location estimation/ It could avoid background noises to be subtracted for enhancing accuracy; Tracking People is difficult, because people can move very fast. Whereas the blind person cannot move as fast as a normal human. We can use the configuration in the current observation and a dynamic model to predict the next observation which has to be synchronized to be equal to the actual measurement. These predictions can then be refined using the Markov chain estimates to Co ordinate the Human Body observation Derived from GPS Tracker

The purpose of Human Tracking for improving accuracy is to provide a better service to tracking people and observing obstacle will be more useful for Security in known and unknown environments, Safety for a normal Human and Blind human was being analyzed according to their walking step variations. This system can be used for tracking normal and blind human against an application/module; the additional features available in our system are steps observation, obstacle detection, user access control, report generators and emergency situation alert.

A human tracking device in which we incorporated Markov Chain algorithm to stably tracks a human with good perception of the distance with high resistance against disturbance and also to use the same device for blind for perceptional differences.

### 1.1 LITERATURE SURVEY

The present day human tracking devices and related techniques have used many number of monitoring systems that monitors intruders. Recently, studies for treating a human image contained in a camera image by image-processing have been very popularly made. Many papers dealt with a method for detecting moving objects using time-lapse of sequential camera images. In some, a human is represented in a particular way with three blob models so as to track him. This concept based on the blob models is based on PFinder developed by MIT Media Lab.[1-2]

#### 1.1.1. Active Badge, Active Bat
Information technology and ubiquitous computing are very important subjects in our era. The advancements in these technologies enable us to transfer data from one point to another in a relatively cheap manner compared to the cost of the technologies used few years earlier.

### 1.1.2 Active Badge
Active badge was originally developed from 1989-1992 at AT&T labs. Active badge makes it possible to locate any person at any given time with an acceptable accuracy. As it was an acceptable accuracy, it uses infrared technology (IR) to transmit data. The advantage of IR solid state emitters is the ability to produce them in a very small and very cheap manner.

### 1.1.3 Active Bat
The idea behind the Active Bat project is to be able to locate a person/badge at any desired position in a place where Active Badges doesn't have much functionality. These devices use ultrasound technology using sonar waves to transmit signals to the receivers that are installed at some intervals on the ceiling.

### 1.1.4 Digital Angel
Digital Angel is a medical/tracking system intended for consumer use by ADS. Originally conceived as a subdermally implanted device, it has combined tracking devices Global Positioning System emitter and an advanced medical monitor that can track heartbeats, blood pressure and other vital signs.

### 1.1.5 Verichip
Verichip is a technology that's gained a lot of attention now days. It is a mall, grain sized RFID chip that is embedded under the skin. RFID tags, is a scanner/emitter system where the scanners emit low-frequency radio wave signals which tells the location coordinate that can be caught by the tags.

### 1.1.9 Cisco Wireless Location Appliance
The Cisco Wireless Location Appliance is the industry's first location solutions that simultaneously track thousands of devices from within the WLAN infrastructure [3-24].

### 1.1.10 Location Tracking Technology
There are numerous location tracking technologies available, under are some of the technologies which work in different concepts.
- ❖ Triangulation does not take into account characteristics such as:
- ❖ Reflection—A wireless signal's reflection off of an object.
- ❖ Attenuation—A physical object's effect on a signal. The coverage circle rings on the triangulation map are must be perfect circles and do not reflect the effect which is imposed from the walls, glass, and other building materials exert on RF. A wall might take away –4 dBm of device signal strength, which affects the accuracy level.

- ❖ Multi-Path—Different paths that an RF signal might take to arrive at a destination, will also cause accuracy deficiency.

### 1.1.11 Blind People Tracking Devices

The days when man used to navigate his way in unfamiliar environments with the help of simple compasses is long gone. [12] Some of the already existing Location Based Service (LBS) providing systems are also been used by the blind persons.

**Drishti,** an Integrated Navigation System for Visually Impaired and Disabled is based on wireless pedestrian navigation system. It integrates many technologies including wearable computers, voice recognition and synthesis, wireless networks, Geographic Information System (GIS) and Global positioning system (GPS).

**Braille Note GPS** is a available talking digital map. Braille Note GPS uses a cell-phone size GPS receiver to relay the location information from GPS satellites. It calculates which is your present location and plots the route to a destination you choose [25-27]

**GPS-GSM Mobile Navigator** combines the GPS's ability to identify location along with the ability of the Global System for Mobile Communications (GSM) to communicate with a base station in a wireless fashion. The navigator is a microcontroller-based system equipped with a GPS receiver and a GSM module. The working principles of the above Location Based Service systems have been utilized in the invention of a route guidance system for the Visually Impaired [28-34].

## 2. THE SYSTEM MODEL
The basic operation of the proposed Human tracking system is that when a Human Movement is detected, a specific sensor-odometer in Human module will produce a signal.

This signal sent from these sensors will trigger the microcontroller, then through the transmitter to location Tracker module.

Through GPS it takes the Longitude and latitude of the current position of the person and passed to the location Tracker Module

The location Tracker module will take the decision according to the Markov Chain Algorithm which is incorporated along with the path guidance system and start sending the information to the base module.

The operation of this tracking system requires certain hardware between the tracker module and the base module.

The Database module will take the Coordinate and store the Information in the Internet.

The operation of the Human tracking system requires certain hardware to activate the motion sensors. The main hardware parts are; motion sensors, sensor driving circuits, GPS, ARM, Transceiver, Buzzer, LCD display if needed and Keypad

Human tracking is the process of reporting and tracking the normal and blind human. Other terminology frequently used to describe this process includes:

1. *Build the chip in human, to be tracked.*
2. *Surveillance the Campus and track where people would be tracked.*
3. *Finding one actual and two alternative route from source to destination (Hi-Tech Block to IT Park).*
4. *Take Longitude and Latitude through GPS Tracker device in traceable route.*
5. *Apply Markov Chain algorithm and normalize the walking steps according to accelerometer readings.*
6. *Comparing the one step distance between Normal and blind human.*
7. *Comparing time taken between blind and normal human.*
8. *Build the comparing Graph for Time and Distance*
9. *Preparing the module for tracking human in surveillance Track.*
10. *Comparing Random walk model with human track Model*
11. *Analyzing Algorithm of Human Tracking Model.*
12. *Finding Obstacle Detection in the surveillance track.*
13. *Build the model for future Activity, based on the type of obstacle.*
14. *Activate the voice tracker as an alert signal for the blind person.*
15. *Upload the tracking path of the blind in the appropriate website.*
16. *Activate the alert message through SMS to the caretaker.*

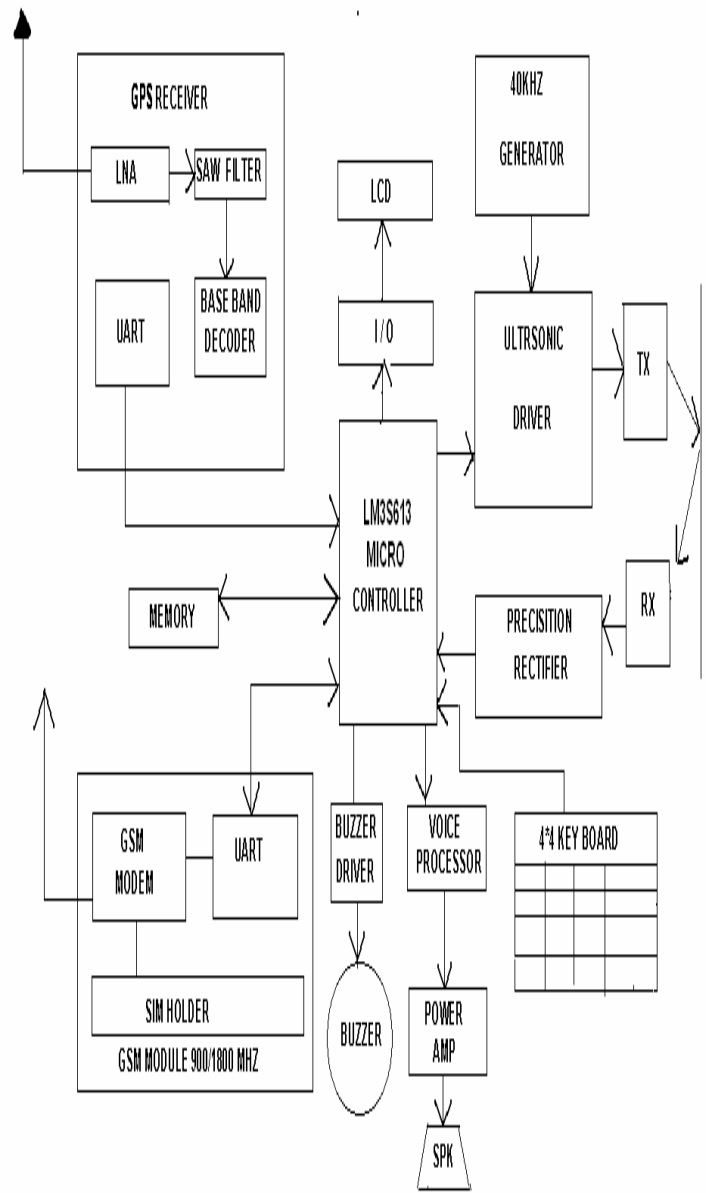

**Figure 1.2 System Design- Diagram**

**2.1 Antenna**

In this apparatus we use a transceiving antenna which receives signal and helps in tracking the human or the blind person.

**2.2 Microcontroller**

The controller has been programmed to acknowledge the received latitude and longitude from receiver and antenna setup and to drive the proper audio track for path guidance.

**2.3 Serial Port**

It is the module for transmitting data bits one by one serially. Serial communication can be used to communicate with computer, GSM Modem, GPS etc. serial communication is perform for long distance communication with cheep cost.

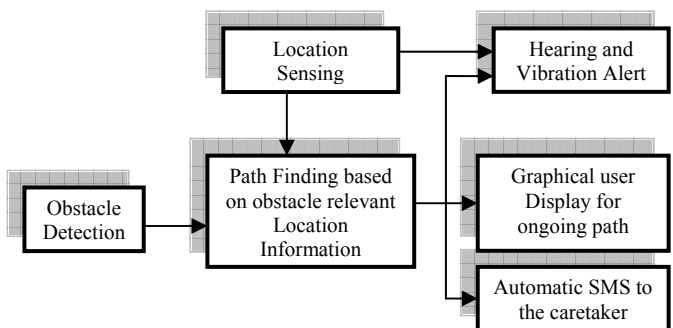

**Figure 1.1 Block Diagram for the system setup**

Serial communication with computer can not be directly performed since computer operates on different voltage logics, so we use RS-232.

### 2.4 Receiver Circuit
It receives the Signal through antenna. It is an interface with microcontroller; the received signal is decoded and displayed on LCD.

### 2.5 The Global Positioning System (GPS)
The Global Positioning System (GPS) is a worldwide radio-navigation system formed from a constellation of 24 satellites and many number of ground stations.

### 2.5.1 A GPS tracking

A GPS tracking device calculates its position by measuring distances between itself and other GPS satellites. In return, the signals emitted by the satellites carry information about the GPS carrier's location.

### 2.5.2 GPS TRACKING MAP

GPS and GIS technology is widely used for determining the position of an object and measuring distance. However, the whole technology is based on maps. Maps taken by satellites and then the application of GIS technology make the whole process successful.

The GIS technology generally uses digital information with the help of digitized data creation methods. By this digitized method a map is converted into a digital format through the use of a computer aided drafting (CAD) program.

Any particular place is marked by its longitudes, latitude and elevation by co-ordinate points. In this way digital satellite images are analyzed and the data in map form represents real world objects with digital form.

### 2.6 Vector Maps:

In this case the data types are prepared by using geometric points, lines, coordinates, polygons etc. for representing any particular object.

### 2.7 Zigbee Modules:

Tracker is a fully web-based system for tracking the people whom we want to track according to their willing. It provides:

- **Tracking** – Tracking the people who are in need.

- **Dataflow** – Automatic routing and notification to the user.

- **Process enforcement** – Managing and enforcing the path to be followed to the destination.

- **Status** – Up-to-the-minute path tracking and future path guidance mechanism for providing safety to the user.

- **Communication** – Capturing distances towards the destination according to on path obstacles.

- **Accountability** – History and make it for future purposes.

The footsteps do not represent the precise location of a person. A Markov chain Monte carlo method is a promising tracking algorithm for pedestrian models [35-36]

### 3. Randomization derived markov chains for the human to move from one stage to another stage

The Markov property, *given the present state,* **future states are independent of the past states**, the description of the present state of location fully captures all the information that could influence the future path prediction evolution of the process. Future states of location will be reached through a probabilistic assumption process instead of a deterministic one, if the path is not a determined one. At each step the system may change its state from the current state to another state, or remain in the same state, pertaining to a certain probability distribution. The probability distribution is proportionated to the identification of the on road obstacles. The changes of location states are called transitions, and the probabilities associated with various state-changes are the used as the predictions for transition probabilities.

The Markov chain is which we have chosen in this paper is a simple random walk where the state spaces is a set of vertices of a normal graph and the transition steps involve moving to any neighbors of the current vertex with equal probability.

Also the Markov chain is a sequence of random variables $X1$, $X_2$, $X_3$, ... with the Markovian Property, given the present state, the future and past states are independent. This is the formula on which the Markov chain works.

$$\Pr\left(X_{n+1} = x \middle| X_n = x_n, \ldots, X_1 = x_1\right) = \Pr\left(X_{n+1} = x \middle| X_n = x_n\right).$$
--(1)

The possible values of $X_i$ form a countable set $S$ is called the state space of the chain.

Markov chains are often described by a directed path, where the edges are labeled by the probabilities which will go from one state to the other states.

- Let Mg be a Markov Chain with states S0…Sn
- The Fundamental Theorem tells us that after a sufficiently large number of time steps, the probability of being in state Si+1 is the same as being in state Si.
- This probability is the probability which tells that if any problem is there on the path, the person has to

stay in the current position without proceeding further.

- This steady-state condition is known as a *stationary distribution*
- The rate at which a Markov Chain converges to a *stationary distribution* is called the *mixing rate r*.
- The mixing rate decides the obstacle occurrence in the forwarding path. Therefore it incorporates the random walk model.

### 3.1 Random Walks

A Random Walk on connected, undirected, non-bipartite Graph G can be modeled as a Markov Chain Mg, where the vertices of the Graph, V(G), are represented by the location states of the Markov Chain and the transition matrix is as follows:

Define the probability of going from location state *i* to location state *j* in *n* time steps as

$$p_{ij}^{(n)} = \Pr(X_n = j \mid X_0 = i) \quad \text{---(2)}$$

and the single-walking step transition as

$$p_{ij} = \Pr(X_1 = j \mid X_0 = i) \quad \text{---(3)}$$

For a time-homogeneous Markov chain:

$$p_{ij}^{(n)} = \Pr(X_{n+k} = j \mid X_k = i) \quad \text{---(4)}$$

and

$$p_{ij} = \Pr(X_{k+1} = j \mid X_k = i). \quad \text{---(5)}$$

so, the *n*-step walking transition satisfies the Chapman-Kolmogororov equation, that for any *k* such that $0 < k < n$,

$$p_{ij}^{(n)} = \sum_{r \in S} p_{ir}^{(k)} p_{rj}^{(n-k)}. \quad \text{---(6)}$$

The marginal distribution $\Pr(X_n = x)$ is the distribution over states at time *n*. The initial distribution is $\Pr(X_0 = x)$. The evolution of the location tracking process through one time step is described by

$$\Pr(X_n = j) = \sum_{r \in S} p_{rj} \Pr(X_{n-1} = r) = \sum_{r \in S} p_{rj}^{(n)} \Pr(X_0 = r). \quad \text{---(7)}$$

A state *j* is said to be **accessible** from a different state *i* (written $i \rightarrow j$) if, given that we are in state *i*, there is a non-zero probability that at definitely some time in the future, the system will be in state *j*. Formally, state *j* is accessible from state *i* if there exists an integer $n \geq 0$ such that

$$\Pr(X_n = j | X_0 = i) > 0. \quad \text{---(8)}$$

Allowing *n* to be zero means that every state is defined to be accessible from itself that is the person is still idle in the current state.

A state *i* is said to communicate between both vertices with state *j* (written $i \leftrightarrow j$) if it is true that both *i* is accessible from *j* and that *j* is accessible from *I, which denotes the source and the destination*. A set of states *C* is a communicating class which comprises that if every pair of states in *C* communicates with each other, and no state in *C* communicates with any state not in C, which is nothing but an equivalence relation. The *c*ommunicating class is closed if the probability of leaving the class is zero, namely that if *i* is in *C* but *j* is not, then *j* is not accessible from *i*.

We have chosen Markovian Property because, a Markov chain is said to be irreducible if its location state space is a communicating class; in an irreducible Markov chain, it is possible to get to any state from any state, so that if the Blind person is missed, it can be possible to be tracked from the history.

As from the previous discussions Mg is a periodic and irreducible, we can apply the Fundamental Theorem of Markov Chains and deduce that Mg converges to a stationary distribution, so that the possibility of occurrence of an obstacle on path is proportional to the time of stationary distribution_mixing rate r

### 3.2 The Lemma:
For all $v \in V$, $\Pi v = d(v) / 2 |E| + r$ ( d(v) = the degree of v)

### 3.3 The Proof
Hitting time (*huv*) – expected number of steps in a Random Walk that starts at *u* and ends upon its first visit to *v*
Commute time (*cuv*) -- expected number of steps in a Random Walk that starts at *u*, visits *v* once and returns to *u*.
(c*uv* = *huv* + *hvu* + *r*)

### 4. Comparison of Poisson process modeling of moving obstacles inter arrival time with respect to the stationary distribution [expected number of walks to avoid on road obstacle]- markov chains: an application for tracking

Let [ y$_n$, n > or = 0 ] be a countable,Markov chain, having state space

S = { 0,1,2,. . . . .} and Transition Probability Matrix..

Let [ N ( t ) , t >or =0] be a Poisson process with rate /.

[y$_n$] and N (t) are independent with each other..

By defining a stochastic process [ x(t) = 1/N(t) > or = 0.

It is a continuous time Markov chain having state space S.

Changes of state of the process [X(t), t > or = 0], take place at the epochs of occurrence of events of the Poisson process [ N(t)].

Thus N(t) denotes the number of transitions by time t.

The mean time spent in a state is the same as the mean inter-occurrence time between two Poisson events.

If mean time spent in a state is /and is the same for all the states.

Here randomization of operational time is done through Poisson events.

Pij(t) of [ X(t) ,t> or = 0] can be completed by conditioning on N[t]

Thus   Pij(t) = Pr{X(t) = j l x (0) = i }

=….. Pr{X(t) = jl X(0) = i , N(t) = n }x

   Pr { N(t) =n l X(0) =i }------          ( 9 )

Now  Pr{N(t) = n l X(0) =i }
       = Pr {N(t) = n } = e yt (yt)$^n$/n!
and Pr { X(t) = j l X(0) = i , N(t) = n }.

**Gives the probability, that the system goes from state i to state j in n transitions.**

So that, Pr { X(t) = j l X(0) = i , N(t) = n } = Pij $^{(n)}$ in terms of n-step transition problem of the chain
[Y$_n$ , n > or = 0 ]
Thus the matrix of transition function of
[ X(t) , t > or = 0] is P(t) = [ Pij$^{(t)}$ ]

Where Pij(t) = e $^{-yt}$ ….Pij$^{(n)}$ (Yt)$^n$ / n!

Let us find the generator of the Q – matrix of
[X(t) , t > or =0].

As h denotes the length of an infinitesimal interval, Then
Pr { N(h) =1}          = yh + och)
Pr { N(h) =0 }         = 1-yh + och)
Pr { N(h) >or = i }     = 0(h),I > or = 2.

ij(n) = … Pij$^{(n)}$ e$^{-yn}$ (yn)$^n$ / n! + 0 (h)

So that
qij = lim Pij(h)/h = yPij exists and is finite,

then
qi = - … qij = -y (1-Pij) =y (Pij-1)
we thus have
Q = Y (P-1)
Thus [X(t),t > or = 0) is a Markov Process ( Continuous time Markov Chain) with the same state process S, and the same initial distribution as [Y$_n$ ,n . or =0] and having as its infinitesimal generator  Q =y(p-1).

When dealing with the incorporation of uncertainty on the data origin in tracking, one should keep in mind that is dealt with multiple-hypothesis tracking.

**4.1 Obstacle Detection**
There are several approaches in the literature for detecting objects using a laser scanner. This region of work focuses on detecting moving and stationary obstacles on the path of pedestrians.

**4.2Efficiency and Accuracy**

Accurate vehicle positioning data and the path accuracy, relative to the walking step distance, is necessary for the collection of a high quality acoustics data set.  Because of the variety of moving vehicle profiles required to be used for the estimation of distances from the pedestrian's walking speed.

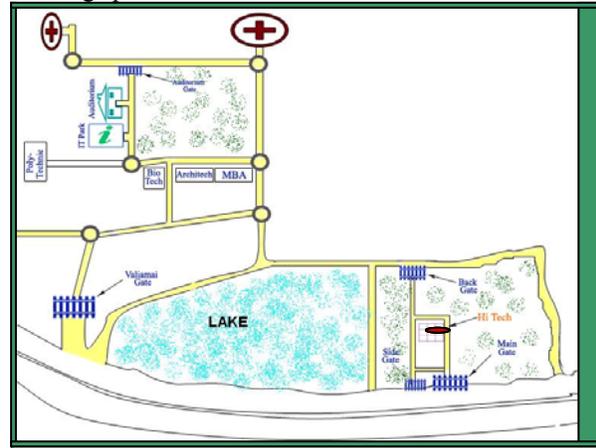

Figure 4.4 Digital Map of the Environment

**4.3 Experimental Values and Analysis:**

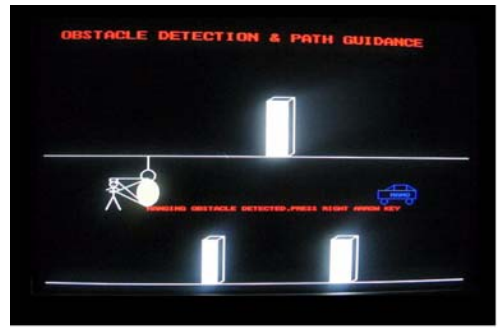

Figure 4.5 A Simulation Environment showing a moving obstacle on the path

Figure 4.4 shows the digital map of the environment,where the results have been taken.Figure 4.5 shows the simulation environment of a moving obstacle on the path of a blind.Before the obstacle crosses the safer distance,the blind have to be intimated about the obstacle.

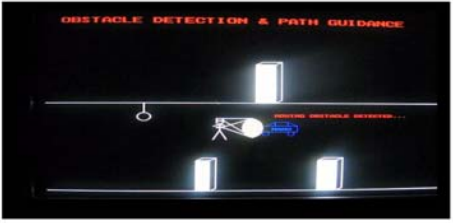

Figure 4.6 A Simulation Environment showing the safer
distance before the moving obstacle on the path

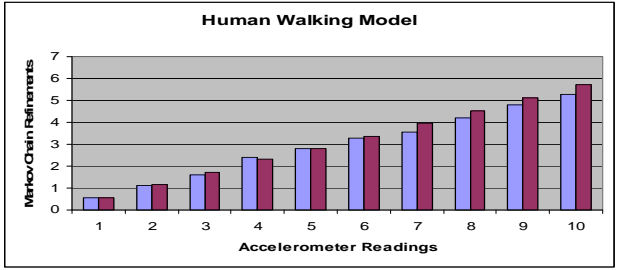

Figure 4.7 The Human Walking model measurements

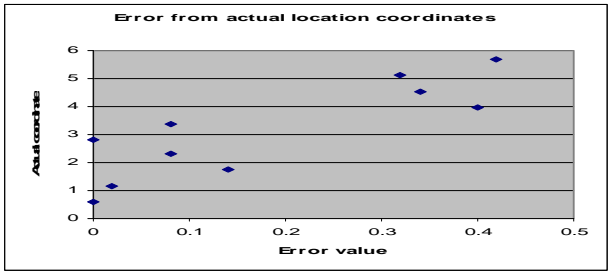

Figure 4.8 The Human Walking model-Error

### 4.4 Implementation

This paper essentially consists of the following major modules: namely, the GPS/GSM interface, Communication system, Route propagation Module and path guidance and Display Module.

### 4.5 GSM-GPS Interface

The positional information received from the GPS receiver is transferred to the microcontroller which is programmed to retrieve the necessary information namely latitude and longitude. This information is directed to the GSM module to be sent as SMS.

### 4.6 Global System for Mobile Communication (GSM) is
a novel communication technology. It is a digital cellular radio network that operates using radio frequency.

### 4.6.1 Advantages of GPS Tracking System

• GPS devices are very small in size and can therefore be installed covertly.

• Monitors the speed and direction of the vehicle.

• Helps to locate your vehicle in case it is stolen or is lost.

• Updates every 60 seconds, or every 30 minutes when vehicle engine is switched off.

### 4.7 Communication System:

The communication system comprises Zigbee module to receive the data of location coordinates and to transmit the location information to the Internet through the GSM interface. An audio tracker is played if there any obstacle has been found on path.

### 4.8 Route Propagation Module:

The route propagation is achieved by incorporating the digital map of the environment and the location data is imposed on the map to identify the current location of the user.

### 4.9 Path Guidance and Display Module:

The distance between the obstacle and the user is found and if it is a non moving or stationary obstacle, the distance is informed to the user by the headphones if he is a blind user. It can also be expressed in terms of visual display.

In this paper we have done an experimental survey of making a normal human and the blind human to walk freely in the known environment and the step difference between the walking step variations and the time difference due to walking step variations. This analysis is useful for devising a new device using Zigbee technology without GPS. As the first prototype model incorporates GPS for locationing; the next models have to be designed without using GPS.

➢ As per our observation one step = 23.2inch, which is equal to 58cm.

➢ It means it is equal to 0.58meter
One normal people's 1 walking step takes time duration equal to 1 sec

And also1 step = .58meter (Calculated)
So, .58 meter distance covered in 1 second
Then, x meter distance Covered in 1/.58 * x Second

➢ Suppose one normal people travel 10 steps then distance covered in  meter is 10 x 0.58 = 5.8meter

➢ As per observation blind take 2.7 sec to cover 0.58 meter

➢ Therefore in One Sec 0.2148 meter is covered

➢ The time taken by blind human being for same distance is 5.8 / 0.2148 = 27sec

For Example: 59.16 meter covered in 1/.58 * 59.16 second = 102second

As per Calculation 0.58 meter distance is covered by Blind people in 2.7second

Then, x meter distance Covered in 2.7/.58 * x Second = 4.655134 * X (Approx 4.66)

=4.66 * x

For Example: 59.16 meter covered in 4.66 * 59.16 second = 275.68 second.

## Distance Graph for Normal Human and Blind People

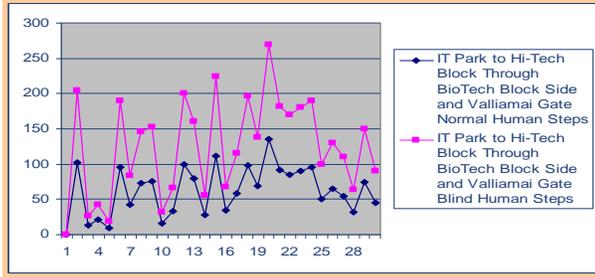

Figure:4.5 IT Park to Hitech Block

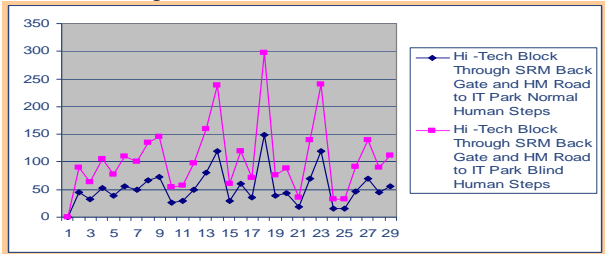

Figure:4.6 Hitech Block to IT Park

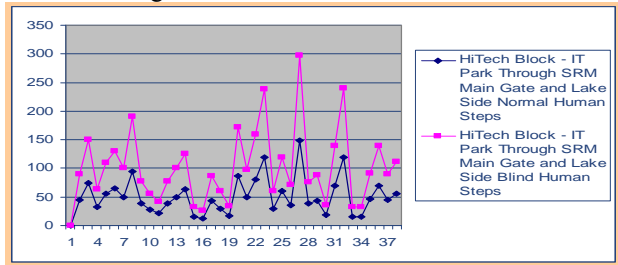

Figure:4.7 Hitech Block to IT Park

## Time Graph in Meter of Human And Blind People

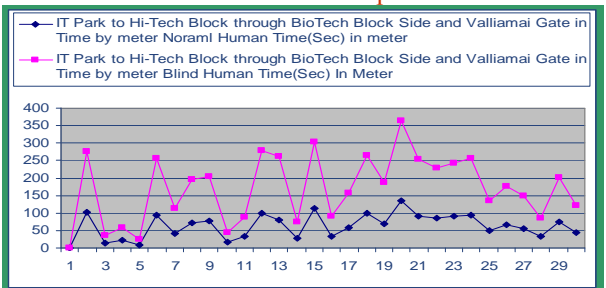

Figure:4.8 Hitech Block to Bio Tech Block

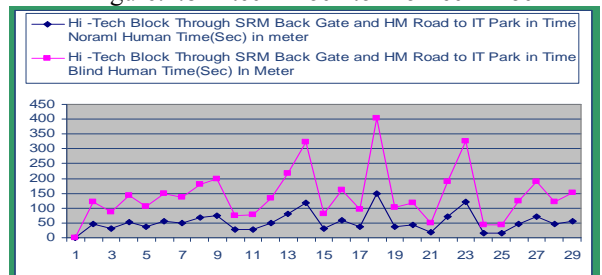

Figure:4.9 Hitech Block to IT Park

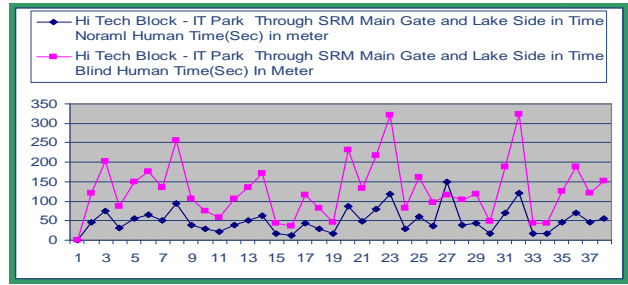

Figure:4.10 Hitech Block to IT Park

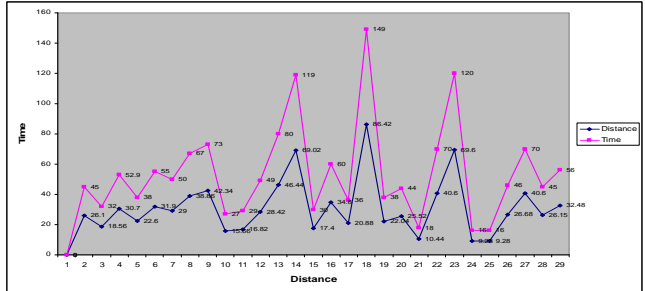

Table 4.1 Time and Distance variation
Figure 4.11 Time and Distance Variation Graph

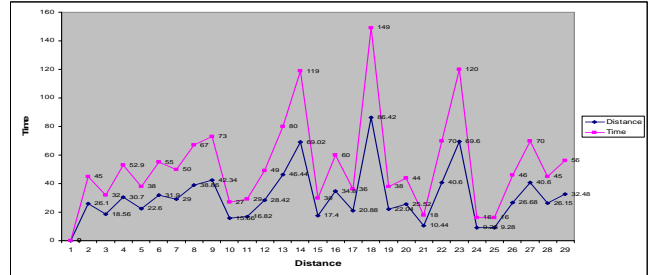

Figure 4.11 Time and Distance variation
For normal human

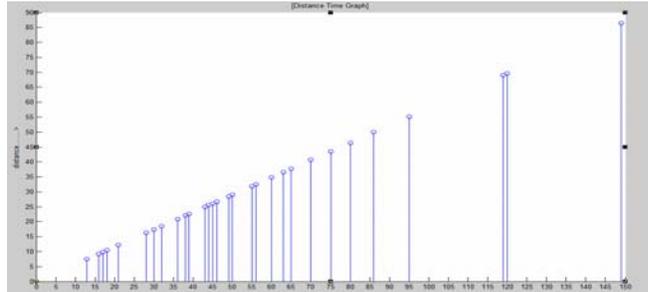

For blind person
Figure 4.12 MatLab Output depicting the Normal and Blind step variations

### 4.10 Result Analysis:

Using Ultrasonic based Sonar for obstacle detection and Zigbee for location sensing from GPS to enhance the location coordinates and the ARM processor for executing the algorithm for Path Finding based on obstacle relevant Location Information. The localization error from actual location coordinates is 0.18 meters. As an improvement the localization error can be reduced by properly fixing the normalization values and applies it in Markov chain algorithm by reducing the noise content in measurements.

| Normal | | Blind | |
|---|---|---|---|
| Distance | Time | Distance | Time |
| 0 | 0 | 0 | 0 |
| 26.1 | 45 | 26.1 | 121.62 |
| 18.56 | 32 | 18.56 | 86.48 |
| 30.7 | 52.9 | 30.7 | 143.06 |
| 22.6 | 38 | 22.6 | 105.31 |
| 31.9 | 55 | 31.9 | 148.65 |
| 29 | 50 | 29 | 135.14 |
| 38.86 | 67 | 38.86 | 181.08 |
| 42.34 | 73 | 42.34 | 197.3 |
| 15.66 | 27 | 15.66 | 72.97 |
| 16.82 | 29 | 16.82 | 78.38 |
| 28.42 | 49 | 28.42 | 132.43 |
| 46.44 | 80 | 46.44 | 216.41 |
| 69.02 | 119 | 69.02 | 321.63 |
| 17.4 | 30 | 17.4 | 81.08 |
| 34.8 | 60 | 34.8 | 162.16 |
| 20.88 | 36 | 20.88 | 97.3 |
| 86.42 | 149 | 86.42 | 402.71 |
| 22.04 | 38 | 22.04 | 102.7 |
| 25.52 | 44 | 25.52 | 118.92 |
| 10.44 | 18 | 10.44 | 48.65 |
| 40.6 | 70 | 40.6 | 189.19 |
| 69.6 | 120 | 69.6 | 324.33 |
| 9.28 | 16 | 9.28 | 43.24 |
| 9.28 | 16 | 9.28 | 43.24 |
| 26.68 | 46 | 26.68 | 124.32 |
| 40.6 | 70 | 40.6 | 189.19 |
| 26.15 | 45 | 26.15 | 121.85 |
| 32.48 | 56 | 32.48 | 151.35 |

**Table 4.1 Normal Human and Blind Person –Distance and Time Values**

**CONCLUSION AND FUTURE WORKS**

This paper presents our work for Normal Human Tracking and Blind Person Tracking based an inexpensive assisted living environment. We have demonstrated how locations of significance can be automatically updated from GPS data. We have also shown the Markov Chain algorithm that can incorporate these locations into a predictive model of the user's movements according to walking step variations. Such location aware computing ensures a higher Accuracy data sampling rate and thus significantly improves the real-time tracking accuracy.

Finally, we present a way to Comparing random walk model and Human Track model for blind people and analyzing the Obstacle detection algorithm in accurate sense of a human being. In the future, we will implement all the proposed methods of Human Track scheduling with blind scheduling mechanisms. We will also implement the system in total Campus test its usefulness in the real-life of the elderly. Although the locationing methods are developed for ultrasonic-based location sensing, the concept of sensing and communication, location awareness, and mobility-consciousness will be useful for many other active sensing systems for the visually impaired.